\documentclass[twocolumn,aps,prl]{revtex4}
\usepackage{epsfig}
\begin{document}
\title{A Quantum Tweezer for Atoms}
\author{Roberto B. Diener$^{1}$, Biao Wu$^{1,3}$, 
Mark G. Raizen$^{1,2}$, and Qian Niu$^{1}$}
\affiliation{$^{1}$Department of Physics, The University of Texas,
Austin, Texas 78712-1081\\
$^{2}$Center for Nonlinear Dynamics, The University of Texas
at Austin, Texas 78712-1081\\
$^{3}$Solid State Division, Oak Ridge National Laboratory,
Oak Ridge, Tennessee 37831-6032}
\date{\today}
\begin{abstract}
We propose a quantum tweezer for extracting a desired number 
of neutral atoms from a reservoir. A trapped Bose-Einstein 
condensate (BEC) is used as the reservoir, taking
advantage of its coherent nature, which can guarantee 
a constant outcome. The tweezer is an attractive quantum dot, 
which may be generated by red-detuned laser light. By moving
at certain speeds, the dot can extract a desired number
of atoms from the BEC through Landau-Zener tunneling. 
The feasibility of our quantum tweezer is demonstrated through
realistic and extensive model calculations.
\end{abstract}
\maketitle

The manipulation and control of isolated single neutral atoms 
has been a long term goal with important applications in quantum 
computing\cite{qbit,chuang} and fundamental physics. Trapping 
and cooling of single neutral atoms was first achieved in magneto-optical 
traps and more recently in a dipole trap
\cite{Kimble,Peng,Kuhr,Grangier}. 
Despite these impressive successes, all existing methods 
share a common weakness: the trapping process itself is random 
and not deterministic. In this letter we propose a
quantum tweezer that can extract a definite number of atoms 
from a reservoir at will, with the atoms in 
the ground state of the tweezer. 
A trapped Bose-Einstein condensate (BEC) is used as a 
reservoir and its coherent nature makes the constancy 
of the output possible. An attractive quantum dot, created by a 
focused beam of red-detuned laser light, serves as a quantum 
tweezer to extract a desired number of atoms from the BEC reservoir.

%%%%%%%%%%%%%%%%%%%%%%%%%%%%%%%%%%%%%%%
\begin{figure}[!htb]
\centering \resizebox *{3in}{2in}
{\includegraphics*{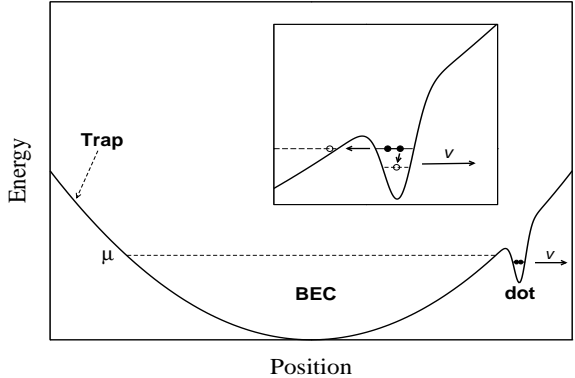}} 
\caption{
A quantum dot (tweezer) moves out of a trapped BEC (reservoir)
with speed of $v$. The inset illustrates that a resonance
occurs as the dot moves further away from the trap center
such that the energy of the atoms 
matches the chemical potential $\mu$ of the condensate. 
If one of the atoms is tunneled  into the BEC, the energy level 
of the dot is lowered, due to the absence of repulsion 
from the lost atom. Thus, no other atom has
a chance of leaking back to the condensate at this
position. 
} 
\label{fig:fig1}
\end{figure}
%%%%%%%%%%%%%%%%%%%%%%%%%%%%%%%%%%%%%%%%

In a typical operation of the quantum tweezer, 
a quantum dot is turned on adiabatically inside the bulk of the BEC 
and moves out of the BEC at a certain speed so that a desired number 
of atoms is extracted (see Figure \ref{fig:fig1}). 
In the initial stage of this operation, it is important that 
the system remains in the ground state of the trap+dot potential.   
The superfluidity of the BEC helps to suppress the excitations 
which might otherwise be induced by the turning on and movement of 
the quantum dot.  The speed of the dot just needs to be slower than 
the speed of sound, and the rate of turning on of the  dot 
potential be smaller than the frequency of phonons whose wavelength 
is comparable to the size of the dot.

The crucial part of the tweezer operation is when the dot moves out of
the BEC.  Inside the BEC, when the coupling between the trap and dot is 
still stronger than the atom self-interaction within the dot,  the system 
is in a coherent state in which the number of atoms in the dot strongly 
fluctuates. 
Outside the BEC, the coupling drops exponentially with 
distance and eventually becomes negligible compared to the 
self-interaction; the eigenstates of the system are then Fock states in 
which 
the dot contains a definite number of atoms.
In the general case, the dot exits the condensate in a superposition of
eigenstates.
However, under certain circumstances, we can 
steer the state into a prescribed final state, with a definite number of 
particles in the dot.

%%%%%%%%%%%%%%%%%%%%%%%%%%%%%%%%%%%%%%%
\begin{figure}[!htb]
\centering \resizebox *{3in}{2in}
{\includegraphics*{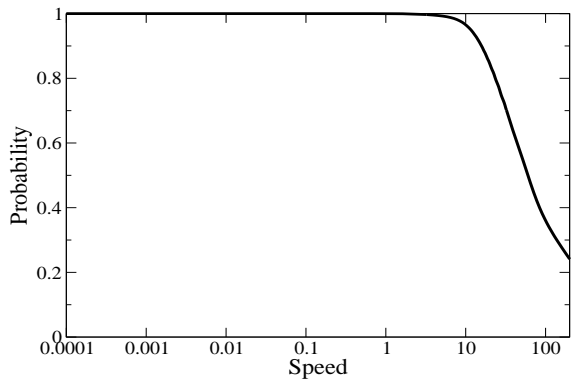}} \caption{The probability of
extracting a single atom as a function of the speed of the dot. 
The calculation was performed for a one dimensional BEC with
$N=10,000$ atoms in a harmonic trap 
with frequency  $\omega=0.005$. The dot is 
a square well with
depth $U_0= 8$ and width $a=1$, and the effective
coupling constant is $g=8$. The units for
all these parameters are defined in the 
text. For sodium, speed is measured in units of 2.75 mm/s, 
so that for many speeds shown it takes a fraction of a second
to extract one atom. 
The plateau exhibited extends several orders of magnitude. 
}
\label{fig:fig2}
\end{figure}
%%%%%%%%%%%%%%%%%%%%%%%%%%%%%%%%%%%%%%%%

Starting from the ground state of the system with the dot at a certain 
position
inside the BEC we start moving the dot outwards.
At
an infinitesimally slow speed, the system always stays in 
the lowest energy state and no atoms are extracted, simply because 
moving out of the BEC costs potential energy of the atoms.   
At some finite speed, the system may get stuck in a non-zero 
number state of the dot, and become de-coupled from the BEC before 
the atoms in the dot have a chance to leak back. In the following, 
we will give a detailed account of this phenomenon through a 
realistic model calculation. In Figure \ref{fig:fig2}, we show a 
result for the probability of extracting a single atom as a function 
of the speed of the dot.  The plateau extends several orders of 
magnitude of the speed, demonstrating the robustness of 
our quantum tweezer.

Our focus is on the crucial stage of the quantum tweezer 
operation--when the dot is leaving the BEC cloud. In this case,
the density is low and the interaction between atoms in the dot 
and atoms in the condensate is weak. 
The state of the system can then be expressed as a combination of atoms
in the dot (with wavefunction $\phi_d$) and atoms in the BEC trap
(with wavefunction $\phi_B$, properly orthogonalized to $\phi_d$
\cite{orthogonalized}).
These 
two wavefunctions are chosen as the adiabatic ground state of the system
when the dot is motionless and the coupling between these two sets of atoms 
is negligible. 

The Hamiltonian of the system is
$
\hat{H} =\int dx \, \hat{\Psi}^\dagger (x) [-{\hbar^2 \over 2M} \nabla^2 +
V_{t}(x)+V_{d}(x,t)
+ {g\over 2}  \hat{\Psi}^\dagger (x)
\hat{\Psi} (x)]\,
\hat{\Psi} (x)$.
We can write $\hat{\Psi}(x)=\phi_B(x) \hat{c} + \phi_d(x) \hat{a}$
in the weak coupling limit, where $\hat{c}$ annihilates an atom in the trap and 
$\hat{a}$ annihilates an atom in the dot.
We shall denote the state with $n$ atoms in the dot (and $N-n$ atoms 
in the BEC) by $|n\rangle$; 
an atom jumping from the dot to the 
BEC corresponds to the transition $|n\rangle \rightarrow |n-1\rangle$. 
Given that $\phi_d(x)$ is much more 
localized than $\phi_B(x)$,
the repulsion felt by the atoms in the dot is stronger than the one 
felt by the ones in the BEC. This asymmetry
between the two potentials yields $n$ much smaller than $N$ in
general and sets our system apart from two-state condensates
discussed elsewhere \cite{Leggett}.

The non-vanishing matrix elements of the Hamiltonian are (for $n\ll N$)
\begin{eqnarray}
\langle n | \hat{H}| n \rangle &=& E_n \equiv n E_1+{n(n-1)\over 2} \nu\,, \\
\langle n | \hat{H} | n+1 \rangle &=& \langle n+1 | \hat{H} | n \rangle = \sqrt{n+1} [\Delta +
n G]\,,\\
\langle n | \hat{H} | n+2 \rangle &=& \langle n+2 | \hat{H} | n \rangle = \sqrt{(n+1)(n+2)}A.
\end{eqnarray}
The parameters depend on the position $x_d$ of the dot
and can be explicitly calculated.
$E_1=\epsilon_d+V_t(x_d)-\mu+4A$ accounts for the energy difference
between the ground state in the dot and the chemical potential
$\mu$ while $\nu=g J_{0,4}$ represents the repulsion an
atom in the dot feels from  another atom there. 
We have defined the generalized overlap integrals 
as $J_{m,n}=\int dx (\phi_B)^m (\phi_d)^n$.
Notice that $E_1$ 
increases as the dot moves away from the center of the BEC.
The off-diagonal
terms are the couplings that allow an atom to tunnel from the dot to
the BEC (or vice versa) either by itself
or in pairs. The two terms in $\Delta = \sqrt{N}[\langle \phi_B |
V_t|\phi_d\rangle +g N   
J_{3,1}]$ correspond to quantum tunneling over a barrier
and the interaction of 
a particle in the dot with three atoms in the BEC trap, respectively; this
last term dominates when the dot is inside the BEC cloud.
Equivalently, $G=g \sqrt{N}J_{1,3}$ is due to the interaction of
three atoms in the dot with one atom in the trap. Finally, 
$A=g NJ_{2,2}/2$ is due to the interaction of two atoms in the trap with
two in the dot. Outside of the BEC the off-diagonal terms vanish exponentially,
since the overlap integrals do so.

Although our scheme works in any dimensionality, we concentrate in
what follows on a dilute, one-dimensional condensate
\cite{Gorlitz} as an example. Such a system can be obtained by
tightly confining the cloud in the transverse directions, in
which the atomic dynamics are frozen out. The coupling constant is
$g=4\pi a_s \hbar/(M_{atom}L^2)$, where $a_s$ is the
$s$-wave scattering length and $L$ is the length of the perpendicular 
confinement.
We shall express our results in the 
following units: length in units of $L_0=1 \mu$m, time in
units of $M_{atom}L_0^2/\hbar$, and energy in units of 
$\hbar^2/(M_{atom}L_0^2)$.
%which we define by $\hbar=M_{atom}=L_0=1$, with $L_0=1 \mum$ 
%being a typical length unit in Bose-Einstein condensed gases.

\begin{figure}[!htb]
\centering \resizebox *{8.0cm}{6cm}
{\includegraphics*{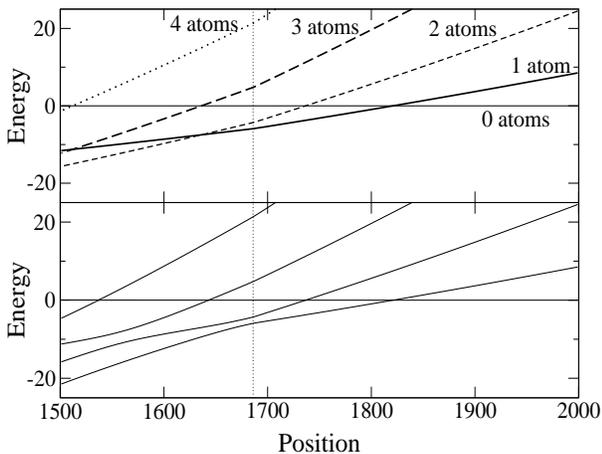}} \caption{
Energy levels as a function of position for different 
numbers of atoms in the dot (bottom panel). In the top
panel the energy levels $E_n$ (ignoring the off-diagonal 
terms) are plotted for comparison. The dotted line 
represents the edge of the condensate. The parameters used
are the ones of Fig. \ref{fig:fig2}.}
\label{fig:fig3}
\end{figure}

We plot in the bottom panel of Figure \ref{fig:fig3} our calculation of 
the energy 
levels as a function of the position for a harmonic trap with 
frequency $\omega=0.005$ and $N=10,000$ atoms. The dot used is a 
square potential with depth $U_0=8$ and width $a=1$;
the coupling constant is $g=8$. 
The edge of the condensate cloud is
marked by the dotted line \cite{change of slope}. 
For comparison, the top panel shows the
curves $E_n(x)$, corresponding to the energies of
states with $n$ atoms in the dot in the absence of the
tunneling terms. The wavefunction
for the BEC was calulated by numerical solution of the Schr\"odinger 
equation in imaginary time \cite{imaginary time solution}.

%%%%%%%%%%%%%%%%%%%%%%%%%%%%%%%%%%%%%%%
%%%%%%%%%%%%%%%%%%%%%%%%%%%%%%%%%%%%%%%%

Let us consider the evolution of the number of atoms in the dot 
as the dot moves out of the BEC, with the help of 
Figure \ref{fig:fig3}. It is possible for an atom to tunnel out 
of the dot
when there is no extra energy required to do so, i.e., when the
energy for $n$ atoms in the dot is equal to the energy of $n-1$
atoms in the dot. This is shown in the top panel of
Figure \ref{fig:fig3} as
the locations where the curves for $E_n$ and $E_{n-1}$ cross. 
These crossings also correspond to the resonance condition 
that we see in Figure \ref{fig:fig1}, the extra energy
due to the $n$-th atom being equal to the chemical potential $\mu$ 
of the condensate \cite{extra}.
The possibility of tunneling out is realized by the off-diagonal terms,
which open up energy gaps in the crossings as seen in the bottom panel of
Fig. 3. As demanded by the quantum adiabatic theorem, starting in the
ground state (the lowest curve) at some position $x$, if the dot moves
infinitesimally slow the system remains in its ground state by losing
one atom at each avoided crossing. When the dot is finally outside the
condensate, no more atoms are left in it. Note that only one atom
is allowed to leak out of the dot at each crossing with the
ground state, due to the
dimishing repulsion between atoms in the dot as there are less atoms in
it (see Fig. 1). The next atom would have a chance of leaking to
the condensate as the dot moves further away from the center of the BEC
trap and lifts up its potential energy. 

On the other hand, if the dot is moving at a finite speed
there is a probability for the system to tunnel through the gap
into an excited state, which corresponds to an atom
not leaking back to the condensate when it is energetically allowed to
do so. In the extreme (sudden) case in which the dot moves at infinite
speed the system remains in its initial state, the
atoms in the dot having no time to leak.
For an atom moving at speed $v$, the probability for
Landau-Zener (LZ) tunneling
\cite{Landau-Zener} on the
width $\delta$ of the gap as $P_{LZ}=\exp(-\delta^2/ 2\alpha v)$,
where $\alpha$ is the difference in the slopes of the two
intersecting curves, which is approximately equal to
$dE_1/dx$.
For a dot moving at fixed speed $v$, the
evolution is adiabatic ($P_{LZ}<0.01$) if $\delta >(9.21 \,
\alpha v)^{1/2}$  and sudden ($P_{LZ}>0.99$) if $\delta<(0.02\,
\alpha v)^{1/2}$. 

We can rewrite the resonance condition as $E_1=-(n-1)\nu$. Using the definition
of $E_1$ and the fact that outside the BEC the off-diagonal terms are
exponentially small, we can see that transitions take place outside the
BEC (i.e. $V(x_d)-\mu>0$) when
\begin{equation}
\epsilon_d+(n-1)\nu <0.
\end{equation}
Notice in particular that the $|1\rangle \rightarrow |0\rangle$ transition
always takes place outside the BEC. Such transitions
are typically
sudden in the sense of the LZ tunneling, 
due to the smallness of $\delta$ there.
We can design a situation in which all transitions with $n\ge n_0$
take place inside the condensate while those with $n<n_0$ occur
outside the cloud. A dot moving at a speed slow enough for all
transitions inside the cloud to be adiabatic extracts then exactly
$n_0$ atoms.

This is demonstrated in Figure \ref{fig:fig2}, where
we show the probability of extracting one atom as a function 
of the velocity of the dot. The calculation was performed assuming 
the system to be initially 
in the ground state some distance inside the BEC and 
integrating the equations of motion with the Hamiltonian matrix (1-3).

%%%%%%%%%%%%%%%%%%%%%%%%%%%%%%%%%%%%%%%
\begin{figure}[!htb]
\centering \resizebox *{8.0cm}{6cm}
{\includegraphics*{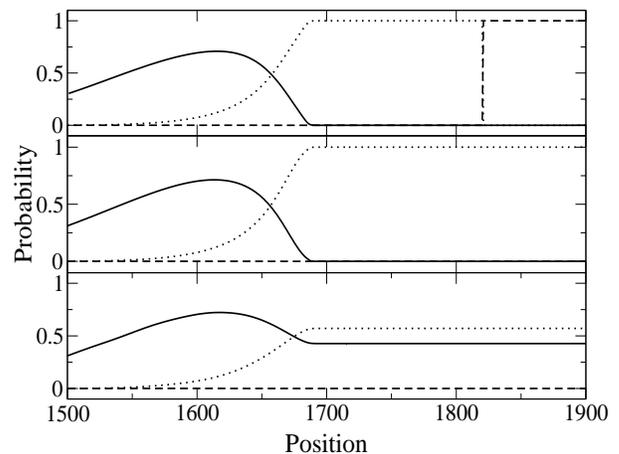}} \caption{
Probability of finding no atoms (dashed), one atom (dotted) and two
atoms (solid line) inside the dot as a function of position for 
three different speeds. The top panel is the adiabatic result
($v\rightarrow 0$), the middle panel is for $v=1$ and the bottom
panel for $v=50$.}
\label{fig:fig4}
\end{figure}
%%%%%%%%%%%%%%%%%%%%%%%%%%%%%%%%%%%%%%%%%%

In Figure \ref{fig:fig4} we show the calculated evolution of the 
probability of finding no atoms (dashed), one atom (dotted),
and two atoms (solid lines) in the dot as a function of position 
for three different speeds of the dot. 
The top panel show the adiabatic result, obtained for
$v$ that is several orders of magnitude smaller than 
the smallest speed in Figure \ref{fig:fig2}. As expected, 
after the $|2\rangle\rightarrow |1\rangle$ transition has taken place 
there is one atom in the dot, while after the 
$|1\rangle\rightarrow |0\rangle$ transition there are none left. The
middle panel corresponds to a slow speed, for which the evolution
is adiabatic everywhere except at the 
$|1\rangle\rightarrow |0\rangle$ transition,
whose LZ tunneling is sudden. Under these circumstances, the dot ends up with
exactly one atom once it is outside the condensate. The last panel
shows the evolution for an even larger speed. In this case the
LZ tunneling takes place only partially and the outcome
corresponds to a superposition of number states. 
One important point illustrated 
clearly in this figure is that there is no definite
number of atoms in the dot during most of the evolution.

\begin{figure}[!htb]
\centering \resizebox *{8cm}{6cm}
{\includegraphics*{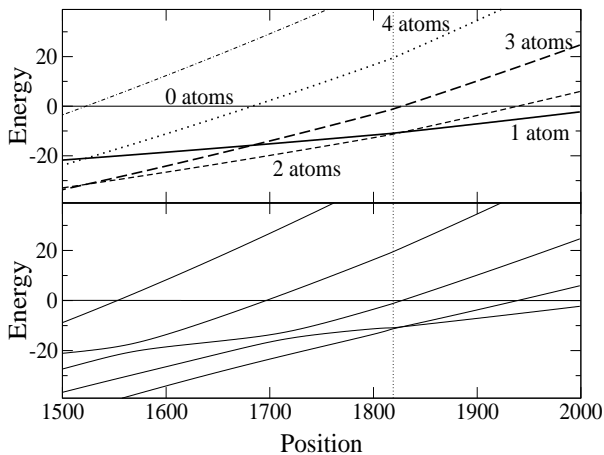}} \caption{
Energy levels as a function of position (bottom panel)
for a one dimensional condensate
with $\omega=0.005$, $N=10,000$ atoms, 
potential depth $U_0 = 13.5$ and width $a= 1$, and effective
coupling constant is $g=10$. The dotted line represents the edge 
of the condensate. The average energy $E_n$ of states 
$|n\rangle$ are plotted in the top panel for comparison.}
\label{fig:fig5}
\end{figure}

\begin{figure}[!htb]
\centering \resizebox *{3in}{2in}
{\includegraphics*{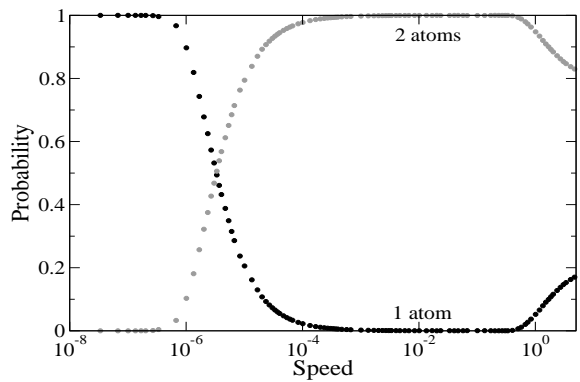}} \caption{The probability of
extracting one (black) and two atoms (gray) as a function of speed for the case 
described in Figure 5. In this case we can see two different plateaus,
showing that the quantized output depends strongly on the speed of the dot.}
\label{fig:fig6}
\end{figure}

We can also find a situation in which for a certain range of
speeds the output is two particles while for a different range (and
for all other parameters fixed) the output is one particle, both
with high certainty. This is achieved by choosing the
$|2\rangle\rightarrow |1\rangle$ transition next to the edge 
of the cloud, so that
both it and the $|3\rangle \rightarrow |2\rangle$ transitions 
have appreciable gaps
opened at the crossing. In Figure \ref{fig:fig5} we
show the energy levels for this case, while in Figure \ref{fig:fig6} 
we see the probability of extracting one and two atoms as a function
of the speed of the dot. We can clearly see that there are two
separate plateaus at different ranges of speed.
%%%%%%%%%%%%%%%%%%%%%%%%%%%%%%%%%%%%%%%
%%%%%%%%%%%%%%%%%%%%%%%%%%%%%%%%%%%%%%%%
%%%%%%%%%%%%%%%%%%%%%%%%%%%%%%%%%%%%%%%
%%%%%%%%%%%%%%%%%%%%%%%%%%%%%%%%%%%%%%%%

Finally, we would like to make some remarks about the generality
of our method. The parameters defining the Hamiltonian matrix 
elements depend on
the density of the trapped BEC at the location of the dot.
We expect the
general behavior found in our one dimensional calculations to
remain true in any number of dimensions, since the motion
of the dot singles out a direction and the other dimensions 
get effectively integrated out.
Moreover, since
the
density of the BEC at the location of the dot is unchanged
by the location of other dots elsewhere, we can consider a train
of dots extracting atoms from the BEC independently of each
other.\

This work has been supported by the NSF and the Welch Foundation.
%\vskip5pt
%\hrule
%\vskip5pt


\begin{thebibliography}{99}
\bibitem{orthogonalized}
Taking $\phi_B^0$ as the ground state solution of the Gross-Pittaevskii
equation without the dot, we define $\phi_B = 
(\phi_B^0 -\beta \phi_d)/\sqrt{1-\beta^2}$. 
Here, 
$\beta$ is the overlap between $\phi_B^0$ and $\phi_d$.
Inside the BEC the solution is approximately equal to the the Thomas-Fermi 
solution
$\phi_B^0(x) = \sqrt{(\mu-V_t(x))/(gN)}$, while outside 
the wavefunction decays exponentially.
\bibitem{change of slope}
The change in the slopes is due to the interaction, through the $A$ 
dependence of $E_1$.
\bibitem{extra} The extra energy due to $n$-th atom in the dot is
$\varepsilon_n=(E_n+n\mu)-(E_{n-1}+\mu(n-1))$, which is represented by the
black level lines in Figure \ref{fig:fig1}. The resonance condition
is $\varepsilon_n=\mu$.
\end{thebibliography}
\end{document}